\documentclass{osa-article}

\journal{osac}

\usepackage{amssymb}
\usepackage{eso-pic}

\articletype{Research Article}

\begin{document}

\title{Degenerate four-wave mixing as a low-power source of squeezed light}

\author{Bonnie L. Schmittberger\authormark{1,2,*}}

\address{\authormark{1}Quantum Technologies Group, The MITRE Corporation, 7515 Colshire Drive, McLean, VA, 22043 USA}

\address{\authormark{2}Approved for Public Release; Distribution Unlimited. Public Release Case Number 20-0216. \copyright 2020 The MITRE Corporation. ALL RIGHTS RESERVED.} 

\email{\authormark{*}bschmittberger@mitre.org} 

\section*{Abstract}
Squeezed light is a quantum resource that can improve the sensitivity of optical measurements. However, existing sources of squeezed light generally require high powers and are not amenable to portability. Here we theoretically investigate an alternative technique for generating squeezing using degenerate four-wave-mixing in atomic vapors. We show that by minimizing excess noise, this technique has the potential to generate measurable squeezing with low powers attainable by a small diode laser. We suggest experimental techniques to reduce excess noise and employ this alternative nonlinear optical process to build a compact, low-power source of squeezed light.

\section{Introduction}
Squeezed states are a useful photonic resource impacting quantum communication, sensing, and metrology. This type of light has unique noise properties in which the noise in one quadrature (e.g., amplitude or phase) can be reduced or ``squeezed'' at the expense of added noise in the other. With the appropriate detection scheme, one can take advantage of the squeezed quadrature to improve the precision of a measurement. For example, squeezed light can be used to improve the phase sensitivity of an interferometer without increasing the optical power~\cite{LIGO1} or reduce the noise of an imaging system~\cite{PhysRevA.95.053849}. Continuous-variable quantum communication and computation can also be enabled using squeezed light, with recent notable advances including a 1.6~km free-space transmission of a squeezed state~\cite{PhysRevLett.113.060502} and the generation of a one-million-mode cluster state~\cite{doi:10.1063/1.4962732}.

Squeezed light is generated using nonlinear optical processes, in which one or more input ``pump'' optical fields interact with a nonlinear material, such as atomic vapors or certain nonlinear crystals, and the mutual light-matter interaction generates optical fields with new frequencies and/or wavevectors. Common methods for generating squeezed states today include non-degenerate four-wave-mixing in atoms and parametric downconversion in crystals. These processes require high input optical powers (up to $\sim$~1 Watt~\cite{PhysRevA.88.043836}) and/or use of a stable cavity in order to enhance the nonlinear interaction~\cite{Schonbeck:18}, both of which inhibit the potential to miniaturize and ruggedize these sources for portable applications. To reduce the size, weight, and power (SWaP) of squeezed light sources, it is crucial to investigate alternative nonlinear optical processes that could enable the production of efficient, low-power, compact sources of squeezed light.


In this paper we theoretically investigate the projected squeezing attainable using a low-power, room-temperature, free-space nonlinear optical process in atomic vapors: degenerate four-wave-mixing. Degenerate four-wave-mixing (DFWM) is appealing as a potential low-SWaP squeezed light source because it is an efficient nonlinear optical process that requires only milliWatt-level powers to generate new, bright optical fields~\cite{PhysRevA.77.013833} and does not require a cavity. Its efficiency is owed to the enhanced light-atom interaction strengths achievable by tuning the frequency of the optical fields close to the atomic resonance. However, this small-detuning regime also generates high rates of spontaneous emission, which gives rise to excess noise that can bury the noise reduction due to squeezing. This excess noise due to spontaneous emission is widely considered to be the culprit that has thus far prevented the observation of squeezing using DFWM~\cite{PhysRevA.32.3803, PhysRevA.34.4929}.

The purpose of the present work is to analyze the potential of using DFWM to generate a low-power source of squeezed light using present practical technologies. To do this, we expand upon previous theoretical models~\cite{PhysRevA.30.343, PhysRevA.31.1622} by incorporating atomic decoherence effects and optical loss. With this expanded model we demonstrate that early DFWM experiments were unsuccessful in measuring squeezing due to high rates of collisional decoherence and/or detection losses rather than spontaneous emission alone. Recent advances in atomic vapor cell technologies and the development of lower-cost, higher quantum efficiency photodiodes enable reduced decoherence rates and detection losses. In addition, recent experiments involving non-degenerate four-wave-mixing have shown that one can generate measurable squeezing even in the presence of strong absorption~\cite{PhysRevA.96.033818} and that excess noise can be reduced by using techniques such as external optical pumping of atoms that have undergone spin-changing collisions with the walls of the vapor cell~\cite{doi:10.1063/1.4980073}. These technological and experimental advances indicate that, by implementing techniques to reduce excess noise, DFWM is a promising technique for observing low-power squeezing in free space.

We will outline our theoretical model for the light-atom interaction in Sec.~\ref{theorysec}. We then analyze two types of degenerate four-wave-mixing: phase-conjugate four-wave-mixing (Sec.~\ref{pcfwmsectheory}) and forward four-wave-mixing (Sec.~\ref{forwardfwm}) . For both beam geometries, we derive the projected squeezing levels as a function of loss and atomic decoherence rates and propose techniques to minimize these sources of excess noise, thus making DFWM a viable platform for generating a low-SWaP source of squeezed light. We present our conclusions in Sec.~\ref{conclusions}.

\section{Theoretical model of the light-atom interaction}
\label{theorysec}

In this section we define our theoretical model of the light-atom interaction by deriving and solving the density matrix equations of motion. This model differs from previous models~\cite{PhysRevA.30.343, PhysRevA.31.1622} in that we include optical loss and atomic decoherence processes.

To model the effects of decoherence, we go beyond the two-level system which is typically used to describe DFWM and consider an empirical four-level model in which atoms can decay to an additional state from which they cannot directly participate in the four-wave-mixing process. We emphasize that this model, detailed below, is not intended to provide a complete physical description of atomic decoherence. Rather, it is used here as a convenient tool through which we can impose an approximate ``decoherence rate'' on the atoms, which reduces the efficiency of the four-wave-mixing process, while still allowing decohered atoms to be optically pumped.

Our model accounts for different types of relaxation processes in the atoms, some of which induce decoherence. Uniform relaxation (decay) is described by the standard spontaneous emission rates in the density matrix equations of motion. Atom-atom and atom-wall collisions can result in an overall phase shift, decay into an unwanted spin state, and/or a rotation of the atomic polarization axis~\cite{PhysRevA.72.023401,PhysRevA.71.012903}. We note that atoms which have undergone a spin-changing collision or some other form of decoherence must be optically pumped before they can participate in a four-wave-mixing process; this in turn reduces the effective atomic density and results in additional spontaneously emitted light.

\begin{figure}
\begin{center}
\includegraphics[scale=0.44]{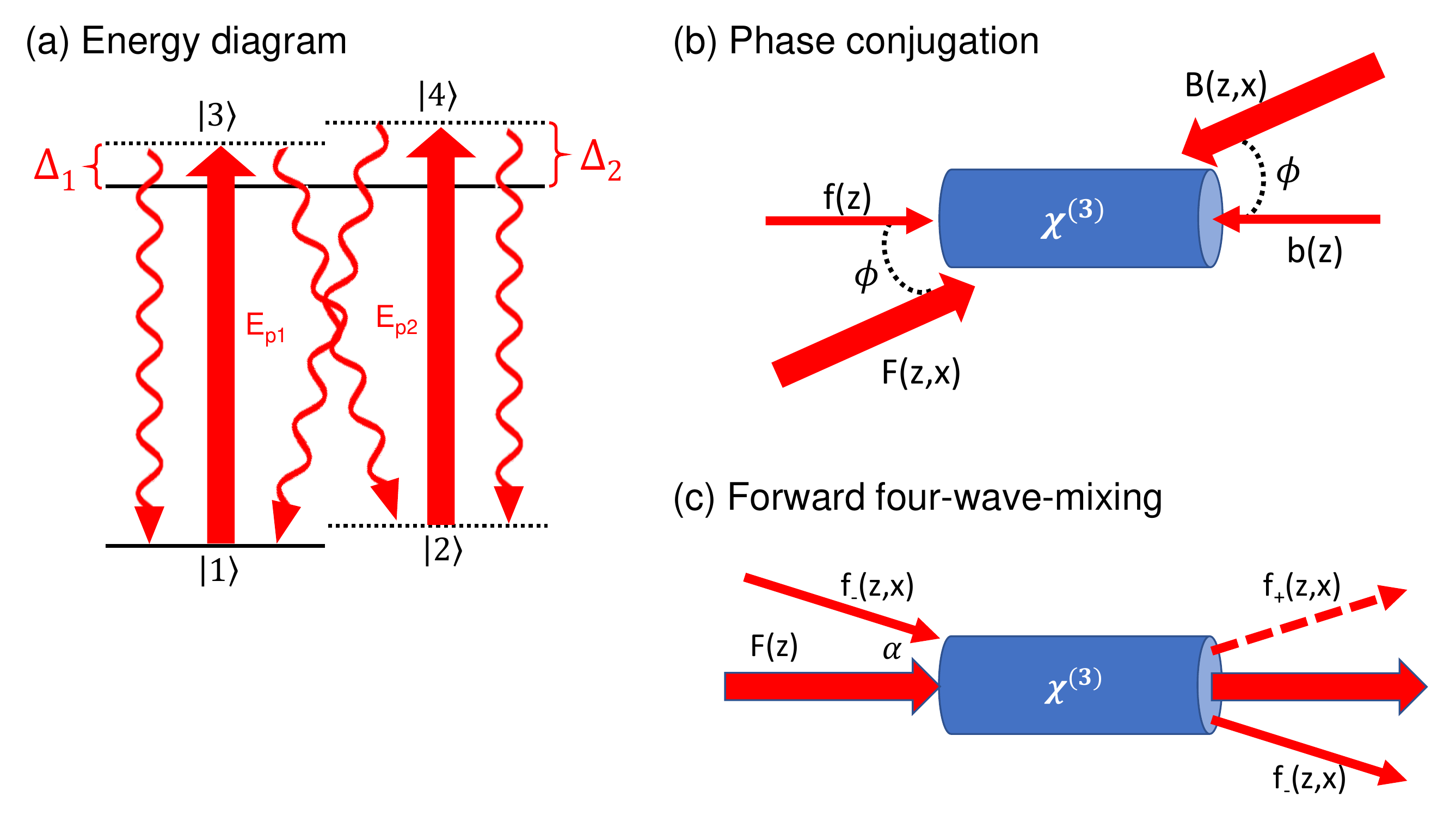}
\caption{(a) The four-level energy scheme considered here. The detunings $\Delta_1$ and $\Delta_2$ define the frequency difference between the pump field and the resonance frequency of the atomic transition. Fields $E_{p1}$ and $E_{p2}$ may in fact be the same pump field depending on the beam geometry, but the subscript is used as an aid to track the field component that is pumping decohered atoms into the desired ground state versus that which is a standard optical pumping cycle. (b) The phase-conjugate/backward four-wave-mixing beam geometry. (c) The forward four-wave-mixing beam geometry.}
\label{energybeamfigure}
\end{center}
\end{figure}

To incorporate the effects of collisional decoherence, we must modify the closed two-level atom description used in previous works to also include additional energy levels into which the atom can decay. We employ the energy level scheme shown in Fig.~\ref{energybeamfigure}(a), where $\left|1\right>$ and $\left|3\right>$ are analogous to the ground and excited states used in a traditional two-level atom description, and level $\left|2\right>$ represents an energy level into which the atoms can decay (such as a separate Zeeman state), but which does not directly participate in the four-wave-mixing process. The atoms must be optically pumped through state $\left|4\right>$ and into state $\left|1\right>$ before they can participate in the DFWM process.

The density matrix equations for a four-level non-degenerate four-wave-mixing process are derived in Ref.~\cite{PhysRevA.88.033845}. We use the same methods here and apply the relevant detuning and decay rate parameters for our degenerate four-wave-mixing system. We define $\Omega_j$ as the Rabi frequency for field $j$, $\Delta_n$ as the optical field detuning from level $n$, and $\Gamma_{nm}$ as the decay rate from level $n$ to level $m$. The coupled density matrix equations for the system shown in Fig.~\ref{energybeamfigure}(a) with density matrix elements $\sigma_{ij}$ in the rotating frame are as follows:
\begin{equation}
\frac{\partial\sigma_{11}}{\partial t}=\frac{i}{2}\left({\Omega}_{p1}^*\sigma_{31}-{\Omega}_{p1}\sigma_{13}\right)+\Gamma_{13}\sigma_{33}+\Gamma_{14}\sigma_{44},
\end{equation}
\begin{equation}
\frac{\partial\sigma_{22}}{\partial t}=\frac{i}{2}\left({\Omega}_{p2}^*\sigma_{42}-{\Omega}_{p2}\sigma_{24}\right)+\Gamma_{23}\sigma_{33}+\Gamma_{24}\sigma_{44},
\end{equation}
\begin{equation}
\frac{\partial\sigma_{33}}{\partial t}=\frac{i}{2}\left({\Omega}_{p1}\sigma_{13}-{\Omega}_{p1}^*\sigma_{31}\right)-\Gamma_{3}\sigma_{33},
\end{equation}
\begin{equation}
\frac{\partial\sigma_{44}}{\partial t}=\frac{i}{2}\left({\Omega}_{p2}\sigma_{24}-{\Omega}_{p2}^*\sigma_{42}\right)-\Gamma_{4}\sigma_{44},
\end{equation}
\begin{equation}
\frac{\partial\sigma_{43}}{\partial t}=\frac{i}{2}\left({\Omega}_{p2}\sigma_{23}-{\Omega}_{p1}^*\sigma_{41}\right)+\left(i\Delta_2-i\Delta_1-\Gamma_{43}\right)\sigma_{43},
\end{equation}
\begin{equation}
\frac{\partial\sigma_{42}}{\partial t}=\frac{i}{2}\left({\Omega}_{p2}\sigma_{22}-{\Omega}_{p2}\sigma_{44}\right)+\left(i\Delta_2-\Gamma_{42}\right)\sigma_{42},
\end{equation}
\begin{equation}
\frac{\partial\sigma_{41}}{\partial t}=\frac{i}{2}\left({\Omega}_{p2}\sigma_{21}-{\Omega}_{p1}\sigma_{43}\right)+\left(i\Delta_2-\Gamma_{41}\right)\sigma_{41},
\end{equation}
\begin{equation}
\frac{\partial\sigma_{32}}{\partial t}=\frac{i}{2}\left({\Omega}_{p1}\sigma_{12}-{\Omega}_{p2}\sigma_{34}\right)+\left(i\Delta_1-\Gamma_{32}\right)\sigma_{32},
\end{equation}
\begin{equation}
\frac{\partial\sigma_{31}}{\partial t}=\frac{i}{2}\left({\Omega}_{p1}\sigma_{11}-{\Omega}_{p1}\sigma_{33}\right)+\left(i\Delta_1-\Gamma_{31}\right)\sigma_{31},
\end{equation}
and
\begin{equation}
\frac{\partial\sigma_{21}}{\partial t}=\frac{i}{2}\left({\Omega}_{p2}^*\sigma_{41}-{\Omega}_{p1}\sigma_{23}\right)+\left(i\Delta_2-i\Delta_1-\Gamma_{21}\right)\sigma_{21}.
\end{equation}
To conserve population, we have ${\partial(\sigma_{11}+\sigma_{22}+\sigma_{33}+\sigma_{44})}/{\partial t}=0$. For succinctness, we define $\tilde{E}_j=E_je^{i\vec{k}_j\cdot\vec{r}}$, $\sigma_{ii,jj}=\sigma_{ii}-\sigma_{jj}$, $\xi_{42}=i\Delta_2-\Gamma_{42}$, and $\xi_{31}=i\Delta_1-\Gamma_{31}$. We also define $d_{31}$ as the transition dipole moment between states $\left|3\right>$ and $\left|1\right>$, $\hbar$ as Planck's constant divded by 2$\pi$, $\epsilon_0$ as the permittivity of free space, and $c$ as the speed of light.

The steady-state solution for the component of the density matrix describing the $\left|3\right>\rightarrow\left|1\right>$ transition is
\begin{equation}
\sigma_{31}=-\frac{i}{\hbar\xi_{31}}d_{31}\tilde{E}_{p1}\sigma_{11,33}.
\end{equation}
The corresponding susceptibility is given by
\begin{equation}
\chi_{31}=-\frac{in_a}{\hbar\epsilon_0\xi_{31}}|d_{31}|^2\sigma_{11,33}.
\label{chi31}
\end{equation}
The intensity of the input pump optical field is defined as $I_p=2\epsilon_0c|E_{p1}|^2$. By solving the coupled density matrix equations to determine the steady state value of $\sigma_{11,33}$, we find that the off-resonant saturation intensity is
\begin{equation}
I_{s\Delta}=\frac{\epsilon_0c\hbar^2\left[\Gamma_{14}\Gamma_{24}\left(\Gamma_{13}+\Gamma_{23}\right)|\xi_{31}|^2+\Gamma_{13}\Gamma_{23}\left(\Gamma_{14}+\Gamma_{24}\right)|\xi_{42}|^2\right]}{|d_{31}|^2\Gamma_{13}\Gamma_{24}\left(\Gamma_{14}+\Gamma_{23}\right)}.
\end{equation}
When $I_p\ll I_{s\Delta}$, the susceptibility in Eq.~\ref{chi31} can be written as the sum of the linear and nonlinear components, $\chi_{31}\approx\chi_{\text{lin}}+\chi_{\text{NL}}$, where
\begin{equation}
\sigma_{11,33}\approx\frac{\Gamma_{14}\Gamma_{24}\left(\Gamma_{13}+\Gamma_{23}\right)|\xi_{31}|^2}{\Gamma_{14}\Gamma_{24}\left(\Gamma_{13}+\Gamma_{23}\right)|\xi_{31}|^2+\Gamma_{13}\Gamma_{23}\left(\Gamma_{14}+\Gamma_{24}\right)|\xi_{42}|^2}\left(1-\frac{I_p}{I_{s\Delta}}\right),
\end{equation}
\begin{equation}
\chi_{\text{lin}}=-\frac{in_a}{\hbar\epsilon_0\xi_{31}}|d_{31}|^2\frac{\Gamma_{14}\Gamma_{24}\left(\Gamma_{13}+\Gamma_{23}\right)|\xi_{31}|^2}{\Gamma_{14}\Gamma_{24}\left(\Gamma_{13}+\Gamma_{23}\right)|\xi_{31}|^2+\Gamma_{13}\Gamma_{23}\left(\Gamma_{14}+\Gamma_{24}\right)|\xi_{42}|^2},
\label{fourlevelchilin}
\end{equation}
and the nonlinear susceptibility is
\begin{equation}
\chi_{\text{NL}}=-\chi_{\text{lin}}\frac{I_p}{I_{s\Delta}}.
\label{chinl}
\end{equation}

We use the parameter $\Gamma_{23}$ to describe the rate at which atoms decay into an unwanted ground state, e.g., due to collisions or other decoherence mechanisms. The experimental values of $\Gamma_{23}$ and $\Gamma_{24}$ are related to the atomic velocity (i.e., the rates at which atoms collide with the walls and pass through the pump beam after collisions, respectively). The average velocity of the atoms is given by $v_{avg}\approx\sqrt{{2k_BT}/{m}}$, where $k_B$ is Boltzmann's constant, $T$ is the atomic temperature, and $m$ is the mass of the atom~\cite{PhysRevA.71.012903}. For a vapor cell temperature of $110~^\circ$C, $v_{avg}\approx308$~m/s for rubidium atoms. For a 1~cm vapor cell size and in the absence of other collisions, atoms would collide with the walls at a rate of approximately $\Gamma_{\text{coll}}\approx2\pi\times62$~kHz~$\approx0.01\Gamma$, where $\Gamma=2\pi\times6$~MHz is the average rate of spontaneous emission in rubidium.

Decoherence rates in atomic vapor cells have been measured to be higher in practice, e.g., $0.2\Gamma$ reported in Ref.~\cite{PhysRevA.88.033845}. This high rate may be due to other inelastic collisions or decoherence mechanisms such as the interaction of atoms with background magnetic fields. Atoms that lose coherence immediately after traversing the pump field will have an even higher decoherence rate (approximately $0.5\Gamma$ for typical beam widths). We therefore expect that a realistic value of $\Gamma_{23}$ will lie in the range of approximately $0.01\Gamma\leq\Gamma_{23}\leq0.5\Gamma$.

We note that $\Gamma_{23}\rightarrow0$ simplifies this framework to that of a two-level atom, i.e., it implies that all population will only undergo optical pumping in the $\left|1\right>\leftrightarrow\left|3\right>$ path. With this assumption, Eq.~\ref{chi31} simplifies to
\begin{equation}
\chi_{_{31}}~\Big|_{\Gamma_{23=0}}=-\frac{n_a|d_{31}|^2\left(\Delta_1/\Gamma_{31}-i\right)}{\hbar\epsilon_0\Gamma_{31}\left(1+\Delta_1^2/\Gamma_{31}^2\right)}\frac{1}{1+|E_{31}|^2/|E_{s\Delta}|^2},
\end{equation}
with $\left|E_{s\Delta}\right|^2={\hbar^2\Gamma_{31}^2\left(1+\Delta_1^2/\Gamma_{31}^2\right)}/\left({4|d_{31}|^2}\right)$, which is the susceptibility of a two-level atom~\cite{boyd2008nonlinear}. 

To summarize our model, previous works on degenerate four-wave-mixing often assume a two-level atom structure where atom number is conserved. Here, we add a separate ground state in order to empirically model loss of atomic population from coherence with the relevant four-wave-mixing transition $\left(\left|1\right>\rightarrow\left|3\right>\right)$. This technique allows us to tune both the rates at which atoms decohere and that for which they are optically pumped back into state $\left|1\right>$ in order to investigate the regimes under which squeezing may occur.

\section{Phase conjugate degenerate four-wave-mixing}
\label{pcfwmsectheory}

We first consider the field geometry shown in Fig.~\ref{energybeamfigure}(b), which is often referred to as backward four-wave-mixing or phase conjugation. We restrict this to the case where the pump-probe angle $\phi$ is large enough such that the only type of four-wave-mixing process that occurs is in the backward (phase-conjugate) geometry. We consider small angles that give rise to forward four-wave-mixing in Sec.~\ref{forwardfwm}. The first step is to solve for the propagation of the pump fields through the atomic vapor in the absence of the weak (generated) fields. The wave equation is
\begin{equation}
\nabla^2\vec{E}-\frac{1}{c^2}\frac{\partial^2\vec{E}}{\partial t^2}=\frac{1}{\epsilon_0c^2}\frac{\partial^2\vec{P}}{\partial t^2}.
\end{equation}

Below threshold for DFWM, the electric field amplitudes of the forward and backward pump beams have exponential solutions $F(z)=\tilde{F}e^{i\delta z}$ and $B(z)=\tilde{B}e^{-i\delta z}$ where $\delta=\frac{k}{2}\chi_{\text{lin}}\left[1-{3I_p}/{I_{s\Delta}}\right]$ and $\tilde{F}$ and $\tilde{B}$ are independent of position.

Above threshold for DFWM, we can define the total electric field $\overrightarrow{\tilde{E}}(t)=\overrightarrow{E}e^{-i\omega t}+\text{c.c.}$, where
\begin{equation}
\overrightarrow{E}={F}(z,x)e^{ik(\text{cos}\theta z+\text{sin}\theta x)}\hat{x}+{B}(z,x)e^{-ik(\text{cos}\theta z+\text{sin}\theta x)}\hat{x}+{f}(z)e^{ikz}\hat{x}+{b}(z)e^{-ikz}\hat{x}.
\end{equation}
Here, all fields are frequency degenerate, $k$ is the wavenumber in vacuum, and we assume $f$ and $b$ are weak fields.

The solution to the wave equation in the case of a single seed beam ($f(0)>0$ and $b(L)=0$) is well-known and was originally derived in Ref.~\cite{Yariv:77}. In the quantized treatment, the solutions for the forward and backward field operators are
\begin{equation}
a_f(z)=\frac{\text{cos}\left[|\kappa|(z-L)\right]}{\text{cos}\left(|\kappa|L\right)}a_f(0)+i\frac{|\kappa|}{\kappa^*}\frac{\text{sin}\left(|\kappa|z\right)}{\text{cos}\left(|\kappa|L\right)}a_b^\dagger(L)
\label{yariv1}
\end{equation}
and
\begin{equation}
a_b^\dagger(z)=\frac{\text{cos}\left(|\kappa|z\right)}{\text{cos}\left(|\kappa|L\right)}a_b^\dagger(L)+i\frac{\kappa^*}{|\kappa|}\frac{\text{sin}\left[|\kappa|(z-L)\right]}{\text{cos}\left(|\kappa|L\right)}a_f(0),
\label{yariv2}
\end{equation}
where
\begin{equation}
\kappa=-\frac{k}{2}\chi_{\text{NL}}.
\end{equation}
The commutation relations are $[a_i,a_j]=0$, $[a_i^\dagger,a_j^\dagger]=0$, and $[a_i,a_j^\dagger]=\delta_{ij}$. The field quadratures are defined as $X_i=\text{Re}[a_i]={a_i+a_i^\dagger}/{2}$ and $Y_i=\text{Im}[a_i]={a_i-a_i^\dagger}/{(2i)}$. We also define the amplitude sum/difference $X_{\pm}=X_f\pm X_b$ and the phase sum/difference $Y_\pm=Y_f\pm Y_b$. We note that for some operator $O$, $\left<\Delta O^2\right>=\left<\left(O-\left<O\right>\right)^2\right>=\left<O^2\right>-\left<O\right>^2$. We take the input fields $a_f(0)$ and $a_b(L)$ to be coherent states, such that $\left<\Delta X_f^2(0)\right>=\left<\Delta Y_f^2(0)\right>=1/4$ and $\left<\Delta X_b^2(L)\right>=\left<\Delta Y_b^2(L)\right>=1/4$. We note that the joint measurement of the fields can be squeezed, but the indivdual quadratures are thermal. For example, the noise of the amplitude quadrature for $a_f$ at $z=L$ is
\begin{equation}
\left<\Delta X_f^2(L)\right>=\frac{1}{4}\left[\text{sec}^2\left(|\kappa|L\right)+\text{tan}^2\left(|\kappa|L\right)\right].
\end{equation}
Therefore, when $|\kappa|L=0$ (i.e., when the nonlinear susceptibility is zero), $\left<\Delta X_f^2(L)\right>=\frac{1}{4}$, and the output is just coherent. For $0<|\kappa|L<\pi/2$, the variance is bigger than that of a coherent state. As $|\kappa|L\rightarrow\pi/2$, the variance approaches $\infty$.

\subsection{Quadrature squeezing}
\begin{figure}
\begin{center}
\includegraphics[scale=0.64]{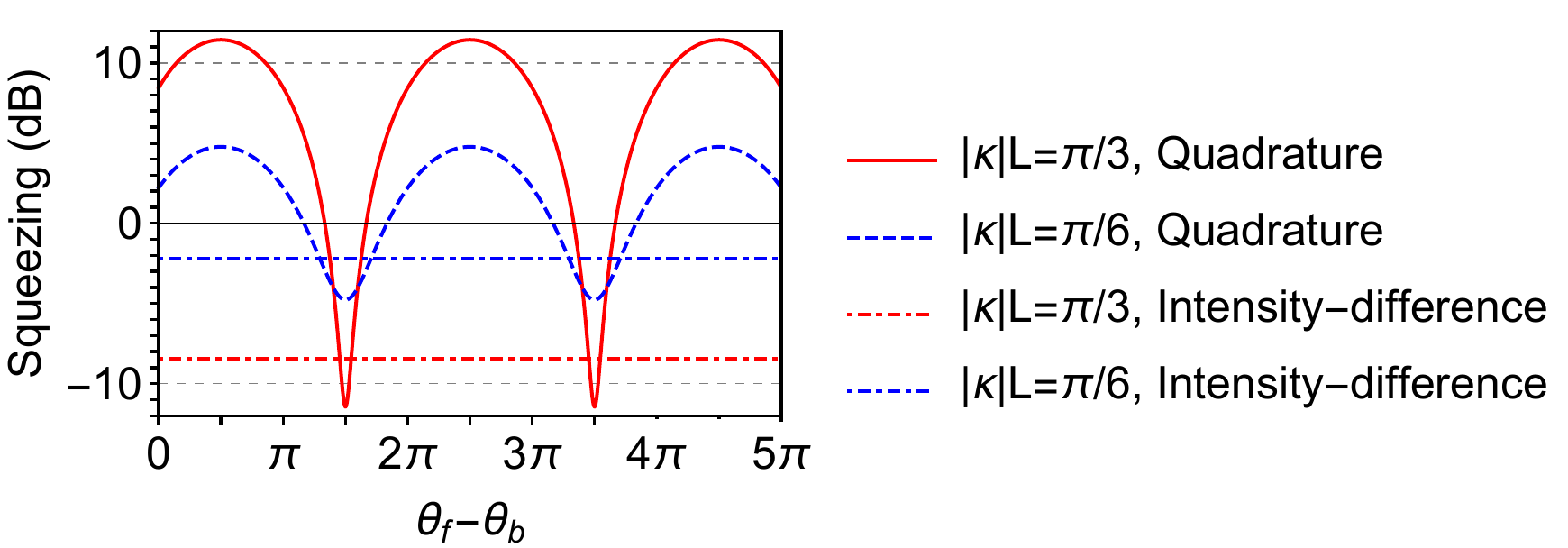}
\caption{The projected quadrature and intensity-difference squeezing generated in the phase-conjugate four-wave-mixing geometry in the ideal case of no loss and with $|\kappa|L=\pi/3$ ($\pi/6$) corresponding to red (blue) curves. The solid and dashed curves correspond to quadrature squeezing and the horizontal dot-dashed lines correspond to intensity-difference squeezing, which is independent of the phase of the measurement. Shot noise corresponds to 0~dB. Negative (positive) values correspond to squeezing (anti-squeezing).}
\label{quadvsintdiff}
\end{center}
\end{figure}

To predict the squeezing associated with a joint quadrature measurement, we calculate the variance associated with the output fields $a_f(L)$ and $a_b(0)$. The quadrature operator for the forward field is $j_f=e^{-i\theta_f}a_f(L)+e^{i\theta_f}a_f^\dagger(L)$ and the quadrature operator for the backward field is $j_b=e^{-i\theta_b}a_b(0)+e^{i\theta_b}a_b^\dagger(0)$, where $\theta_f$ and $\theta_b$ are the phases of the homodyne detectors. The joint quadrature operator is $j=j_f+j_b$. In the ideal case of no loss, the variance is
\begin{equation}
\left<\Delta j^2\right>=2\left|\text{sec}\left(|\kappa|L\right)+ie^{-i(\theta_f-\theta_b)}\frac{\kappa}{|\kappa|}\text{tan}\left(|\kappa|L\right)\right|^2.
\label{quad1noloss}
\end{equation}
We note that in this ideal case, the noise is independent of the input field strength $a_f(0)$; i.e., any noise on the seed beam is present in each generated beam and then canceled by the nature of the joint measurement. The measurable quadrature squeezing in units of decibels (dB) is given by
\begin{equation}
M_{\text{Q}}=10~\text{log}_{10}\left[\frac{\Delta^2j}{\Delta^2j_{\text{SN}}}\right],
\label{quadsqueezing}
\end{equation}
where $j_{\text{SN}}$ corresponds to the noise that would be measured using coherent fields. For homodyne detection, where the local oscillator power dominates the measurement, we can take shot noise of the joint homodyne measurement to be the noise in the case where $|\kappa|L\rightarrow0$, i.e., where there is no four-wave-mixing~\cite{PhysRevA.95.063843}. The results of Eqs.~\ref{quad1noloss} and~\ref{quadsqueezing} are shown in Fig.~\ref{quadvsintdiff} for example values of $|\kappa|L$.

Here, higher values of $|\kappa|L$ correspond to stronger light-atom interactions and hence more squeezing. In practice, one can increase $|\kappa|L$ by increasing the atomic density, applied optical intensity, or the length of the vapor cell, for example. In the limit where $|\kappa|L\rightarrow\pi/2$ and at $\theta_f-\theta_b=3\pi/2$, $M_{\text{Q}}\rightarrow-\infty$. In other words, in this ideal case of no loss, there is an operating point for which we expect ``infinite squeezing.'' Of course, in practice, there always exists some loss, which quickly reduces the squeezing to finite values.

\begin{figure}
\begin{center}
\includegraphics[scale=0.45]{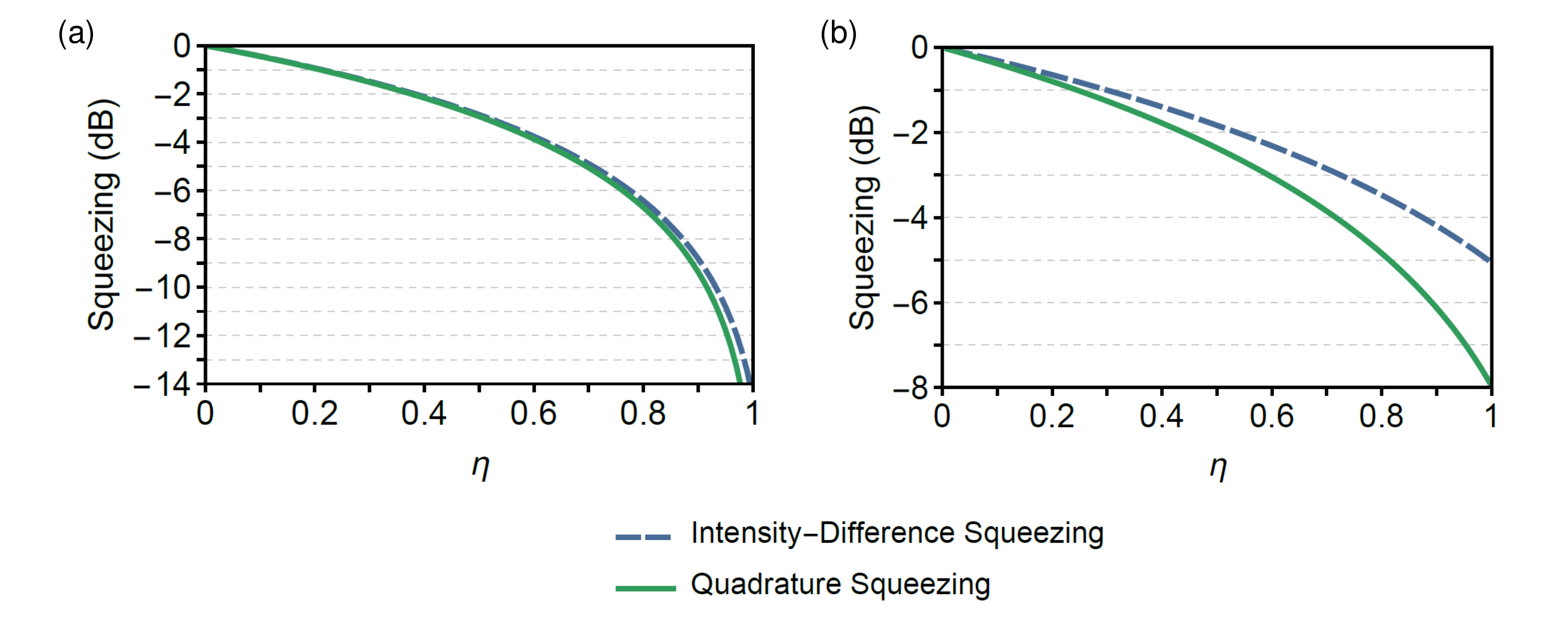}
\caption{Squeezing as a function of $\eta$ for (a) the phase-conjugate beam geometry and (b) the forward beam geometry. For both figures, we assume the following parameters: $\Gamma_{23}=0.1\Gamma$, $\Gamma=2\pi\times6$~MHz, $\Gamma_{24}=2\pi\times30$~kHz (approximately the rate at which an atom will traverse the vapor cell), $d_{31}=1.1\times10^{-29}~\text{C}\cdot\text{m}$, $\Delta_2\approx\Delta_1=\Delta=50\Gamma$, $\text{L}=3~\text{cm}$, $I_p=8~\text{W/cm}^2$, $n_a=10^{16}~\text{atoms/m}^3$.}
\label{squeezingvsloss}
\end{center}
\end{figure}

To account for loss, we use the procedure outlined in the Appendix. To model the effects of Doppler broadening, we must modify the detuning according to $\Delta-\vec{k}\cdot\vec{v}$, where $\vec{k}$ is the optical wavevector and $\vec{v}$ is the atomic velocity~\cite{PhysRevA.55.R1601,PhysRevA.49.1326}. For a thermal distribution, the atomic motion is described by the Maxwell-Boltzmann distribution $S(v_z)\propto e^{-v_z^2/u^2}$, where $u=\sqrt{2k_BT/m}$ is the average velocity of the atoms and $v_z$ is the projection of the atomic velocty along $\hat{z}$, i.e., the optical axis. To determine the Doppler-broadened nonlinear coefficient, we integrate the density matrix elements over all velocities. The projected squeezing including loss is shown in Fig.~\ref{squeezingvsloss}(a) for typical experimental parameters, which follows the expected trend that squeezing is optimized for no loss, and all squeezing is lost for 100$\%$ loss.

The projected squeezing as a function of input pump intensity in the case of $30\%$ loss (transmittion parameter $\eta=0.7$) is shown in Fig.~\ref{plotsummaries}(a). Here, we use typical experimental parameters for atomic density, field detuning, and pump intensity, as defined in the caption. The various curves correspond to different decoherence rates, where the best squeezing is obtained for the slowest rates of decoherence. The oscillatory solutions in Eqs.~\ref{yariv1} and~\ref{yariv2} give rise to oscillatory squeezing results, which are not immediately intuitive. (We note that the curves representing faster decoherence rates also oscillate, but their first minima occur at higher pump intensities.) In this phase-conjugate beam geometry, the counterpropagating pump beams generate a sinusoidal interference pattern across the atoms, which in turn creates a sinusoidal intensity-depedent index of refraction. One can therefore interpret this process as a type of ``nonlinear'' Bragg scattering, where the efficiency of scattering depends on the index of refraction and the angle of the optical field relative to the spatially varying index of refraction. In the forward four-wave-mixing beam geometry described in Sec.~\ref{forwardfwm}, there is only a single pump beam, and the projected squeezing correspondingly has a non-oscillatory solution.

\subsection{Intensity-difference squeezing}
It is also worthwhile to analyze the projected squeezing of an alternative detection scheme, known as ``intensity-difference detection,'' which is generally simpler to implement than homodyne detection in practice. In this case, one sends each beam generated by the four-wave-mixing process to one of two photodiodes on a balanced detector. This type of direct measurement that does not require a local oscillator is considered to be particularly useful for simplifying the detection apparatus for certain applications in quantum imaging and metrology.

To model intensity-difference squeezing, we consider the number operators of the output fields, $n_f=a_f^\dagger(L)a_f(L)$ and $n_b=a_b^\dagger(0)a_b(0)$.
The intensity difference is $N_-=n_f-n_b$. The noise of the intensity-difference measurement is therefore $\left<N_-\right>=\gamma_f-\gamma_b$, where $\gamma_{f}$ ($\gamma_b$) is the seed photon number for the forward (backward) mode. In the vacuum-seeded case, $\gamma_f=\gamma_b=0$. For the seeded case considered below, I assume the forward mode is seeded with a weak beam of photon number $\gamma_f=\gamma\gg1$ and $\gamma_b=0$.

The intensity difference squeezing level in dB is given by
\begin{equation}
M_{\text{ID}}=-10~\text{log}_{10}\left[\frac{\Delta^2N_-}{\Delta^2\hat{n}_{\text{SN}}}\right],
\label{squeezing}
\end{equation}
where $\Delta^2\hat{n}_{\text{SN}}$ is the shot noise, i.e., the variance of the measurement of two coherent beams having the same intensities as the forward and backward beams generated in the four-wave-mixing process.

To calculate shot noise, we need to first calculate the number of photons in each beam at the output of the four-wave-mixing process, i.e., $\left<a_f^\dagger(L)a_f(L)\right>$ and $\left<a_b^\dagger(0)a_b(0)\right>$. For the coherent beams used for our shot noise measurement, the variance is simply equal to the average value, and thus $\left<\Delta\hat{n}_{\text{SN}}^2\right>=\left<\hat{n}_f\right>+\left<\hat{n}_b\right>$. In the single-seeded case,
\begin{equation}
\left<\Delta\hat{n}_{\text{SN}}^2\right>=\gamma\text{sec}^2\left(|\kappa|L\right)+(\gamma+2)\text{tan}^2\left(|\kappa|L\right).
\end{equation}
Comparing the noise $\left<\Delta N_-^2\right>=\gamma$ to the shot noise, one can see that the four-wave-mixing process has not altered the average fluctuations on the output beams. However, it has increased the photon number while keeping the noise unchanged, hence giving rise to a squeezing of
\begin{equation}
{M_{\text{SQ,PC}, \eta=1}=-10~\text{log}_{10}\left[\frac{\gamma}{\gamma\text{sec}^2\left(|\kappa|L\right)+(\gamma+2)\text{tan}^2\left(|\kappa|L\right)}\right]}.
\end{equation}
In the limit where $|\kappa|L\rightarrow\pi/2$, the trend approaches infinite squeezing in this ideal case of no loss. This limiting behavior agrees with the results of Ref.~\cite{PhysRevA.31.1622}. The expected intensity-difference squeezing for example values of $|\kappa|L$ is shown in Fig.~\ref{quadvsintdiff}. It is interesting to note that the expected intensity-difference squeezing is less than that expected for the optimal quadrature measurement. This result can be attributed to the fact that the intensity-difference measurement essentially neglects all phase information, and hence performs a measurement related to a projection of the amplitude quadrature rather than the full noise ellipse. This result is consistent with squeezing generated by non-degenerate four-wave-mixing~\cite{PhysRevA.95.063843}.


The intensity-difference squeezing in the presence of loss in the case where $\eta_f=\eta_b=\eta$ and $\gamma\gg1$ is given by
\begin{equation}
M_{\text{ID,PC}}=10\text{log}_{10}\left[1-\eta\left(2+\frac{4}{\text{cos}\left(2|\kappa|L\right)-3}\right)\right].
\end{equation}
The projected squeezing as a function of the transmission parameter $\eta$ is shown in Fig.~\ref{squeezingvsloss}(a) for example experimental parameters defined in the caption. The projected squeezing level follows closely to the projected quadrature squeezing for these particular parameters where the light-atom interaction is strong. It is worthwhile to note that $|\kappa|L<\pi/2$ is below the threshold for spontaneous degenerate four-wave-mixing, but we still expect to observe squeezing because the process is seeded by a weak probe beam.

\begin{figure}
\begin{center}
\includegraphics[scale=0.45]{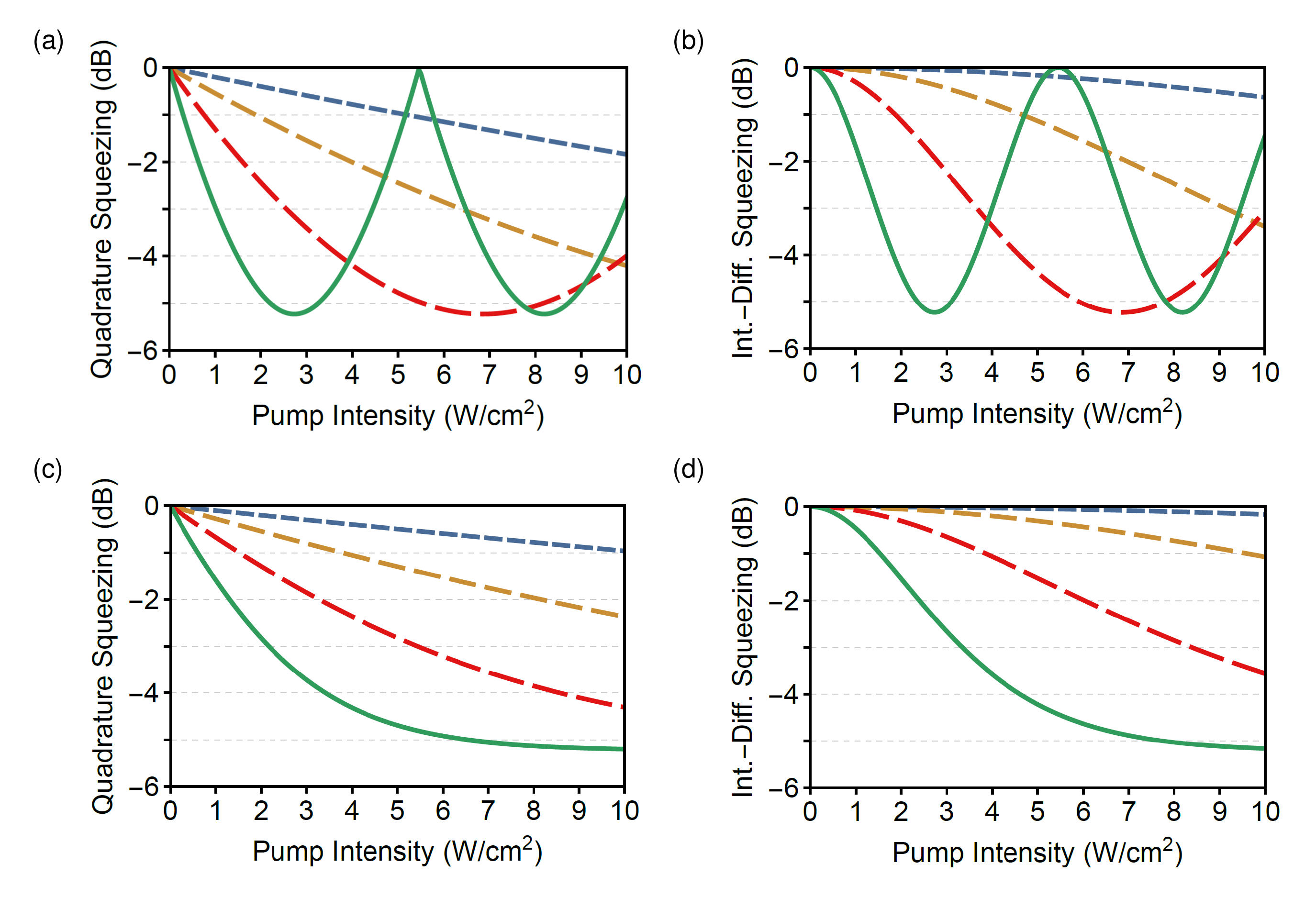}
\caption{Quadrature and intensity-difference squeezing for the (a,b) phase-conjugate and (c,d) forward four-wave-mixing beam geometries as a function of pump intensity for the example case of $\eta=0.7$. In order from the shortest to longest dashing, the curves show the cases $\Gamma_{23}=0.5\Gamma$, $\Gamma_{23}=0.2\Gamma$, $\Gamma_{23}=0.1\Gamma$, and $\Gamma_{23}=0.05\Gamma$, with the latter being the solid curve. All other parameters are the same as that for Fig.~\ref{squeezingvsloss}.}
\label{plotsummaries}
\end{center}
\end{figure}

The intensity-difference squeezing for phase-conjugate four-wave-mixing process as a function of pump intensity is shown in Fig.~\ref{plotsummaries}(b). For high rates of decoherence ($\Gamma_{23}>0.5\Gamma$), one requires fairly high pump intensities to observe measurable squeezing. However, for sufficiently low decoherence rates, we project that more than 3 dB of squeezing is attainable at low pump intensities (less than 10~W/cm$^2$).

Phase-conjugate DFWM is therefore a promising technique for generating squeezing using low optical powers. We note that, in practice, this process requires careful alignment of the counterpropagating pump fields to ensure optimal gain in the four-wave-mixing process, where gain is defined as the ratio of the power in the output probe beam to the input seed beam. In addition, the present work employs a single-mode plane wave description of the optical fields, whereas in practice, multimode emission can occur~\cite{Dawes672}. In the high-gain regime, multimode emission can give rise to cross-correlations among the generated fields which potentially suppress squeezing in a given spatial mode. We therefore anticipate that it will be beneficial to work with fairly low gains in practice to suppress multimode four-wave-mixing. The phase-conjugate four-wave-mixing process also requires a sufficiently small angle between the pump and probe beams to maximize their overlap and hence the effective length $L$, but it should be kept sufficiently large to suppress additional four-wave-mixing processes.

\section{Forward four-wave mixing}
\label{forwardfwm}
In the case of very small pump-probe angles, another type of four-wave-mixing process can occur. Unlike the phase conjugation process, the forward four-wave-mixing process shown in Fig.~\ref{energybeamfigure}(c) is only nearly phase-matched, but it can also give rise to high gain under certain conditions.

The electric field above threshold for forward four-wave-mixing is
\begin{equation}
\overrightarrow{E}=\overrightarrow{F}(z)e^{ikz}+\overrightarrow{f}_-(z,x)e^{ik(\text{cos}\theta z-\text{sin}\theta x)}+\overrightarrow{f}_+(z,x)e^{ik(\text{cos}\theta z+\text{sin}\theta x)},
\end{equation}
where $k$ is the wavenumber in vacuum, and $f_+$ and $f_-$ are weak fields. In this case, with $f_\pm(z,x)=f_\pm^\prime e^{i\delta_kz}$, $\delta_k=(\delta+k-k\text{cos}\alpha)$, and operating with the pump-probe angle $\alpha\approx\sqrt{\chi_{\text{lin}}I_p/I_{s\Delta}}$ (typically 3-5~mrad) for optimizing the phase-matching condition, we find the solutions to the wave equation for the probe and conjugate fields in the quantized treatment are
\begin{equation}
a_-(z)=i\frac{\nu}{|\nu|}\text{sinh}(|\nu|z)a_+^{\dagger}(0)+\text{cosh}(|\nu|z)a_-(0)
\end{equation}
and
\begin{equation}
a_+^{\dagger}(z)=\text{cosh}(|\nu|z)a_+^{\dagger}(0)-i\frac{\nu^*}{|\nu|}\text{sinh}(|\nu|z)a_-(0),
\end{equation}
where
\begin{equation}
\nu=-\frac{k}{2\text{cos}\theta}\chi_{\text{lin}}\frac{I_p}{I_{s\Delta}}.
\end{equation}
These solutions agree with the results of Ref.~\cite{PhysRevA.30.1568}.

Using the same calculation methods presented above for phase-conjugate four-wave-mixing, the quadrature noise in the ideal case of no loss for forward four-wave-mixing goes as
\begin{equation}
{\left<\Delta j^2\right>=2\left|-i\text{cosh}(|\nu|L)+e^{-i(\theta_++\theta_-)}\frac{\nu}{|\nu|}\text{sinh}(|\nu|L)\right|^2},
\end{equation}
where $\theta_{\pm}$ represent the homodyne detector phases for the $f_\pm$ fields. The projected quadrature squeezing is shown in Fig.~\ref{squeezingvsloss}(b) as a function of loss and in Fig.~\ref{plotsummaries}(c) as a function of pump intensity. The squeezing shows the expected trend, where it is optimized for low loss and higher pump intensities. At low intensities, we find that the forward four-wave-mixing case is not projected to produce as much squeezing as the phase-conjugate beam geometry for these parameters, but it can still potentially produce more than 3 dB of squeezing with less than 10~W/cm$^2$ of pump intensity for a decoherence rate of $\Gamma_{23}=0.1\Gamma$.

For the intensity-difference measurement, the noise in the ideal, no loss, single-seeded case is given by $\left<\Delta N_-^2\right>=\gamma$, which is the same as that for the phase-conjugate four-wave-mixing beam geometry. The shot noise is
\begin{equation}
\left<\Delta\hat{n}_{\text{SN}}^2\right>=-1+(1+\gamma)\text{cosh}(2|\nu|L).
\end{equation}
Hence, the intensity-difference squeezing for this forward geometry in the limit where $\gamma\gg1$ goes as
\begin{equation}
M_{\text{ID,FFWM}}=10\text{log}_{10}\left[\frac{1}{\text{cosh}(2|\nu|L)}\right].
\end{equation}
In the presence of loss where $\eta=\eta_+=\eta_-$,
\begin{equation}
M_{\text{ID,FFWM}}=10\text{log}_{10}\left[1-\eta+\eta\text{sech}(2|\nu|L)\right].
\end{equation}
The intensity-difference squeezing as a function of $\eta$ is shown in Fig.~\ref{squeezingvsloss}(b) for example experimental parameters defined in the caption. We also show the projected squeezing as a function of pump intensity in Fig.~\ref{plotsummaries}(d). We find that for both quadrature and intensity-difference measurements in the forward four-wave-mixing case, the nonlinearity saturates above some pump intensity (just above 10~W/cm$^2$ for $\Gamma_{23}=0.01\Gamma$). For pump intensities above this saturated value, the squeezing does not substantially increase, and hence it is not necessary to work well above the saturation intensity.

\section{Conclusions and Outlook}
\label{conclusions}
We have shown that degenerate four-wave-mixing in the phase-conjugate and forward beam geometries can produce more than 3 dB of squeezing for low ($\lesssim10$~W/cm$^2$) pump intensities for sufficiently low loss and long coherence times. To understand why squeezing using DFWM has not yet been observed experimentally, it is necessary to examine typical values of loss ($1-\eta$) and decoherence rates ($\Gamma_{23}$). Loss arises from the vapor cell windows, optics, the quantum efficiency of the detectors, and, in the case of homodyne detection, the visibility of the fringes generating by interfering the generated field with a local oscillator. Typically, AR-coated optics will only give rise to a few percent loss, largely due to polarization imperfections. The glass wall at the exit of an AR-coated cell should have $<1\%$ loss, but in practice can have more and be detuning-dependent due to atoms coating the cell walls. It is critical to heat the vapor cells in such a way that the coldest part of the cell does not coincide with the optical paths. One of the largest sources of loss can arise from the photodiode in the detection apparatus, and it is critical to work with the highest possible quantum efficiency detector to minimize loss.

We hypothesize that previous experimental works were unable to observe squeezing using degenerate four-wave-mixing due to a combination of optical loss and high rates of decoherence, especially from wall collisions. Additional sources of noise in experiments may also arise from asymmetries in the optical fields and imperfect alignment along with nonlinear lensing effects of the pump beam(s) that can generate multimode coupling. By incorporating techniques for reducing decoherence and minimizing optical loss, we predict that squeezing is attainable using degenerate four-wave-mixing even with low ($\approx10$~mW) optical powers. Techniques for reducing decoherence include the use of OTS-coated cells or a separate pump beam for optically pumping atoms that have collided with the walls of the vapor cell~\cite{doi:10.1063/1.4980073}. These techniques allow atoms to immediately participate in four-wave-mixing with the primary pump beam, which reduces spontaneous emission from optical pumping in the spatial modes of interest. With the more recent advances in vapor cell quality and lower-cost high-quantum-efficiency detectors, we expect that degenerate four-wave-mixing can enable the development of an efficient, low-power source of squeezed light.

\section*{Acknowledgements}
We thank Dr. Brielle Anderson for many helpful discussions in recent years on squeezing and noise. We gratefully acknowledge funding from the MITRE Innovation Program.

\noindent{\copyright 2020 The MITRE Corporation. ALL RIGHTS RESERVED.}

\bibliography{squeezingbib}

\begin{thebibliography}{10}
\newcommand{\enquote}[1]{``#1''}

\bibitem{LIGO1}
{LIGO Collaboration}, \enquote{Enhanced sensitivity of the ligo gravitational
  wave detector by using squeezed states of light,}
  {\protect\JournalTitle{Nature Photonics}} \textbf{7}, 613 EP -- (2013).

\bibitem{PhysRevA.95.053849}
A.~Kumar, H.~Nunley, and A.~M. Marino, \enquote{Observation of spatial quantum
  correlations in the macroscopic regime,} {\protect\JournalTitle{Phys. Rev.
  A}} \textbf{95}, 053849 (2017).

\bibitem{PhysRevLett.113.060502}
C.~Peuntinger, B.~Heim, C.~R. M\"uller, C.~Gabriel, C.~Marquardt, and
  G.~Leuchs, \enquote{Distribution of squeezed states through an atmospheric
  channel,} {\protect\JournalTitle{Phys. Rev. Lett.}} \textbf{113}, 060502
  (2014).

\bibitem{doi:10.1063/1.4962732}
J.-i. Yoshikawa, S.~Yokoyama, T.~Kaji, C.~Sornphiphatphong, Y.~Shiozawa,
  K.~Makino, and A.~Furusawa, \enquote{Invited article: Generation of
  one-million-mode continuous-variable cluster state by unlimited time-domain
  multiplexing,} {\protect\JournalTitle{APL Photonics}} \textbf{1}, 060801
  (2016).

\bibitem{PhysRevA.88.043836}
N.~V. Corzo, Q.~Glorieux, A.~M. Marino, J.~B. Clark, R.~T. Glasser, and P.~D.
  Lett, \enquote{Rotation of the noise ellipse for squeezed vacuum light
  generated via four-wave mixing,} {\protect\JournalTitle{Phys. Rev. A}}
  \textbf{88}, 043836 (2013).

\bibitem{Schonbeck:18}
A.~Sch\"{o}nbeck, F.~Thies, and R.~Schnabel, \enquote{13 db squeezed vacuum
  states at 1550 nm from 12 mw external pump power at 775 nm,}
  {\protect\JournalTitle{Opt. Lett.}} \textbf{43}, 110--113 (2018).

\bibitem{PhysRevA.77.013833}
A.~M.~C. Dawes, L.~Illing, J.~A. Greenberg, and D.~J. Gauthier,
  \enquote{All-optical switching with transverse optical patterns,}
  {\protect\JournalTitle{Phys. Rev. A}} \textbf{77}, 013833 (2008).

\bibitem{PhysRevA.32.3803}
M.~W. Maeda, P.~Kumar, and J.~H. Shapiro, \enquote{Observation of optical
  phase-sensitive noise on a light beam transmitted through sodium vapor,}
  {\protect\JournalTitle{Phys. Rev. A}} \textbf{32}, 3803--3806 (1985).

\bibitem{PhysRevA.34.4929}
M.~D. Reid and D.~F. Walls, \enquote{Quantum theory of nondegenerate four-wave
  mixing,} {\protect\JournalTitle{Phys. Rev. A}} \textbf{34}, 4929--4955
  (1986).

\bibitem{PhysRevA.30.343}
R.~S. Bondurant, P.~Kumar, J.~H. Shapiro, and M.~Maeda, \enquote{Degenerate
  four-wave mixing as a possible source of squeezed-state light,}
  {\protect\JournalTitle{Phys. Rev. A}} \textbf{30}, 343--353 (1984).

\bibitem{PhysRevA.31.1622}
M.~D. Reid and D.~F. Walls, \enquote{Generation of squeezed states via
  degenerate four-wave mixing,} {\protect\JournalTitle{Phys. Rev. A}}
  \textbf{31}, 1622--1635 (1985).

\bibitem{PhysRevA.96.033818}
J.~D. Swaim and R.~T. Glasser, \enquote{Squeezed-twin-beam generation in
  strongly absorbing media,} {\protect\JournalTitle{Phys. Rev. A}} \textbf{96},
  033818 (2017).

\bibitem{doi:10.1063/1.4980073}
L.~Zhu, X.~Guo, C.~Shu, H.~Jeong, and S.~Du, \enquote{Bright narrowband
  biphoton generation from a hot rubidium atomic vapor cell,}
  {\protect\JournalTitle{Applied Physics Letters}} \textbf{110}, 161101 (2017).

\bibitem{PhysRevA.72.023401}
M.~T. Graf, D.~F. Kimball, S.~M. Rochester, K.~Kerner, C.~Wong, D.~Budker,
  E.~B. Alexandrov, M.~V. Balabas, and V.~V. Yashchuk, \enquote{Relaxation of
  atomic polarization in paraffin-coated cesium vapor cells,}
  {\protect\JournalTitle{Phys. Rev. A}} \textbf{72}, 023401 (2005).

\bibitem{PhysRevA.71.012903}
D.~Budker, L.~Hollberg, D.~F. Kimball, J.~Kitching, S.~Pustelny, and V.~V.
  Yashchuk, \enquote{Microwave transitions and nonlinear magneto-optical
  rotation in anti-relaxation-coated cells,} {\protect\JournalTitle{Phys. Rev.
  A}} \textbf{71}, 012903 (2005).

\bibitem{PhysRevA.88.033845}
M.~T. Turnbull, P.~G. Petrov, C.~S. Embrey, A.~M. Marino, and V.~Boyer,
  \enquote{Role of the phase-matching condition in nondegenerate four-wave
  mixing in hot vapors for the generation of squeezed states of light,}
  {\protect\JournalTitle{Phys. Rev. A}} \textbf{88}, 033845 (2013).

\bibitem{boyd2008nonlinear}
R.~Boyd and D.~Prato, \emph{Nonlinear Optics}, Nonlinear Optics Series
  (Elsevier Science, 2008).

\bibitem{Yariv:77}
A.~Yariv and D.~M. Pepper, \enquote{Amplified reflection, phase conjugation,
  and oscillation in degenerate four-wave mixing,} {\protect\JournalTitle{Opt.
  Lett.}} \textbf{1}, 16--18 (1977).

\bibitem{PhysRevA.95.063843}
B.~E. Anderson, B.~L. Schmittberger, P.~Gupta, K.~M. Jones, and P.~D. Lett,
  \enquote{Optimal phase measurements with bright- and vacuum-seeded su(1,1)
  interferometers,} {\protect\JournalTitle{Phys. Rev. A}} \textbf{95}, 063843
  (2017).

\bibitem{PhysRevA.55.R1601}
W.~J. Brown, J.~R. Gardner, D.~J. Gauthier, and R.~Vilaseca,
  \enquote{Amplification of laser beams propagating through a collectionof
  strongly driven, doppler-broadened two-level atoms,}
  {\protect\JournalTitle{Phys. Rev. A}} \textbf{55}, R1601--R1604 (1997).

\bibitem{PhysRevA.49.1326}
M.~Pinard, R.~W. Boyd, and G.~Grynberg, \enquote{Third-order nonlinear optical
  response resulting from optical pumping: Effects of atomic motion,}
  {\protect\JournalTitle{Phys. Rev. A}} \textbf{49}, 1326--1336 (1994).

\bibitem{Dawes672}
A.~M.~C. Dawes, L.~Illing, S.~M. Clark, and D.~J. Gauthier,
  \enquote{All-optical switching in rubidium vapor,}
  {\protect\JournalTitle{Science}} \textbf{308}, 672--674 (2005).

\bibitem{PhysRevA.30.1568}
P.~Kumar and J.~H. Shapiro, \enquote{Squeezed-state generation via forward
  degenerate four-wave mixing,} {\protect\JournalTitle{Phys. Rev. A}}
  \textbf{30}, 1568--1571 (1984).

\bibitem{Li:17}
T.~Li, B.~E. Anderson, T.~Horrom, B.~L. Schmittberger, K.~M. Jones, and P.~D.
  Lett, \enquote{Improved measurement of two-mode quantum correlations using a
  phase-sensitive amplifier,} {\protect\JournalTitle{Opt. Express}}
  \textbf{25}, 21301--21311 (2017).

\end{thebibliography}

\newpage
\pagebreak
\clearpage

\section*{Appendix: Optical loss}
\begin{figure}
\begin{center}
\includegraphics[scale=0.22]{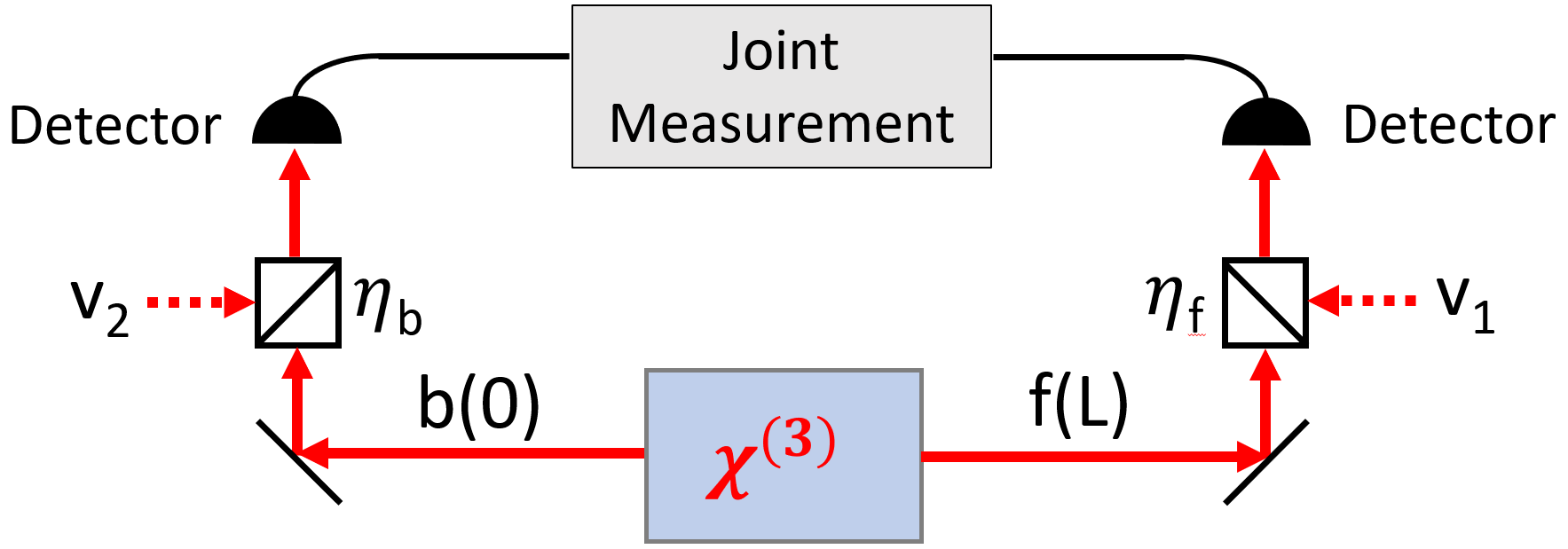}
\caption{A schematic of the theoretical model for incorporating loss that models beam splitters of transmission $\eta_{f,b}$, which also introduce vacuum noise. The vacuum fields are represented by the field operators $v_{1,2}$.}
\label{lossschematic}
\end{center}
\end{figure}

We model loss using the same method presented in Ref.~\cite{Li:17}, in which ``beam splitters" of transmission $\eta_{f,b}$ are inserted into the path of the probe and conjugate beams, as shown in Fig.~\ref{lossschematic}. Here we describe the case of the phase-conjugate four-wave-mixing beam geometry, but it is straightforward to apply this procedure to the forward four-wave-mixing case. We define the initial optical state before the four-wave-mixing process as
\begin{equation}
v_0=
\begin{pmatrix}
a_f(0)\\
a_f^\dagger(0)\\
a_b(L)\\
a_b^\dagger(L)
\end{pmatrix}.
\end{equation}
The output states are defined as
\begin{equation}
\begin{pmatrix}
a_f(L)\\
a_f^\dagger(L)\\
a_b(0)\\
a_b^\dagger(0)
\end{pmatrix}=\mathcal{M}\cdot v_0,
\end{equation}
where
\begin{equation}
\mathcal{M}=
\begin{pmatrix}
\text{sec}\left(|\kappa|L\right) & 0 & 0 & i\frac{|\kappa|}{\kappa^*}\text{tan}\left(|\kappa|L\right)\\
0 & \text{sec}\left(|\kappa|L\right) & -i\frac{|\kappa|}{\kappa}\text{tan}\left(|\kappa|L\right) & 0\\
0 & i\frac{\kappa}{|\kappa|}\text{tan}\left(|\kappa|L\right) & \text{sec}\left(|\kappa|L\right) & 0\\
-i\frac{\kappa^*}{|\kappa|}\text{tan}\left(|\kappa|L\right) & 0 & 0 & \text{sec}\left(|\kappa|L\right)\\
\end{pmatrix}.
\end{equation}
The transmissions of the beam splitters model linear optical loss, such as reflections off surfaces and imperfect detector quantum efficiencies. These beam splitters add vacuum noise, as shown in Fig.~\ref{lossschematic} by the vacuum states $v_1$ and $v_2$. We define
\begin{equation}
\mathcal{L}=
\begin{pmatrix}
\sqrt{\eta_f} & 0 & 0 & 0\\
0 & \sqrt{\eta_f} & 0 & 0\\
0 & 0 & \sqrt{\eta_b} & 0\\
0 & 0 & 0 & \sqrt{\eta_b}
\end{pmatrix},
\end{equation}
which describes the transmission through the beam splitters, and the vector
\begin{equation}
\mathcal{V}=
\begin{pmatrix}
i\sqrt{1-\eta_f}v_1\\
-i\sqrt{1-\eta_f}v_1^\dagger\\
i\sqrt{1-\eta_f}v_2\\
-i\sqrt{1-\eta_f}v_2^\dagger\\
\end{pmatrix},
\end{equation}
which describes the coupling of vacuum noise into the optical fields. Then the final state is
\begin{equation}
v_{\text{fin}}=
\begin{pmatrix}
a_f\\
a_f^\dagger\\
a_b\\
a_b^\dagger
\end{pmatrix}=
\mathcal{L}\cdot\left(\mathcal{M}\cdot v_0\right)+\mathcal{V}.
\end{equation}

For the phase-conjugate four-wave-mixing geometry, the squeezing as a function of the transmission parameter $\eta$ is shown in Fig.~\ref{squeezingvsloss} in the case $\eta_f=\eta_b=\eta$. Squeezing is impaired for increasing loss, and no squeezing is present for $100\%$ loss.

\end{document}